\documentstyle[12pt,cite]{article} 
\begin{document}

~\hspace*{5.0cm} ADP-AT-98-2\\
~\hspace*{5.9cm} {\it Astropart. Phys.}, submitted\\

\begin{center}
{\large \bf Cut-offs and pile-ups in shock acceleration spectra}\\[1cm]
R.J. Protheroe$^a$, Todor Stanev$^b$\\$^a$Department of Physics and
Mathematical Physics\\ The University of Adelaide, Adelaide, Australia
5005\\$^b$Bartol Research Institute, University of Delaware, 
Newark, DE 19716, USA\\[2cm]
\end{center}

\begin{center}
\underline{Abstract}\\
\end{center}

We have examined cutoffs and pile-ups due to various processes in
the spectra of particles produced by shock acceleration, and
found that, even in the absence of energy losses, the shape of
the spectrum of accelerated particles at energies well below the
nominal maximum energy depends strongly on the energy dependence
of the diffusion coefficient.  This has implications in many
areas, for example, in fitting the observed cosmic ray spectrum
with models based on power-law source spectra and rigidity
dependent diffusive escape from the galaxy.  With continuous
energy losses, prominent pile-ups may arise, and these should be
included when modelling synchrotron X-ray and inverse Compton
gamma-ray spectra from a shock-accelerated electron population.

We have developed a Monte Carlo/numerical technique to model the
shape of the spectrum for the case of non-continuous energy
losses such as inverse Compton scattering in the Klein-Nishina
regime.  We find that the shapes of the resulting cut-offs differ
substantially from those arising from continuous processes, and
we suggest that such differences could be observable through their
effect on the spectrum of radiation emitted by a population of
recently accelerated electrons as, for example, may exist in
young supernova remnants.\\

\noindent {\bf PACS codes/keywords:} 96.40 Cosmic rays; 96.40.De
Composition, energy spectra, and interactions; 98.70.Sa Cosmic
rays (including sources, origin, acceleration, and interactions)

\newpage

\section{Introduction}

Cosmic rays with energies up to 100 TeV are thought to arise
predominantly through shock acceleration by supernova remnants
(SNR) in our Galaxy\cite{Lag83}.  A fraction of the cosmic rays
accelerated should interact within the supernova remnant and
produce gamma--rays
\cite{DruryAharonianVolk,GaisserProtheroeStanev96}.
%
  The EGRET instrument on the Compton Gamma Ray Observatory has
 observed gamma ray signals above 100 MeV associated with at least
 two supernova remnants -- IC~443 and $\gamma$~Cygni~\cite{Esposito96}.
 (It is however possible that the gamma ray emission from IC~443 is
 associated with a pulsar within the remnant rather than the remnant
 itself~\cite{Brazier96}).
 Further evidence for acceleration in SNR comes from the recent
 ASCA observation of two spatially resolved regions with non-thermal
 X--ray emission from  SN~1006\cite{Koyama95}. These non-thermal
 X-rays are interpreted as synchrotron emission by electrons
 accelerated in the remnant up to energies as high as 100
 TeV~\cite{Reynolds96,Mastichiadis96}. Donea and
 Biermann\cite{DonauBiermann96} however suggest it may be bremsstrahlung
 from much lower energy electrons. The spectrum of synchrotron
and inverse Compton emission will depend on the the electron
spectrum.  As we shall demonstrate in this paper the effect of a
cut-off in the spectrum due to escape losses, energy losses or
interactions can considerably modify the shape of the
electron spectrum over several decades in energy below the nominal
cut-off energy.

In this paper, we adopt a simple leaky-box model of shock
acceleration based on the physical picture of acceleration at
plane non-relativistic shocks.  This simple model will then be
used to explore the effect of cut-offs due to the finite size of
the accelerator and energy-loss and interaction processes. 

\section{Leaky-box picture of shock acceleration}

The leaky-box picture of shock acceleration~\cite{SzaboProtheroe}
is based on a simple heuristic treatment of Fermi acceleration.
See \cite{ProtheroeCR} for a simple treatment based on that in Gaisser's
book\cite{Gaisser}.  More detailed and rigorous treatments are
given in several review
articles\cite{Drury,BlandfordEichler87,BerezhkoKrymsky88,JonesEllison91}.

The rate of gain of energy of relativistic particles at
non-relativistic shocks is given by
\begin{equation}
 {dE \over dt}  = {\langle \Delta E \rangle \over t_{\rm cycle}}
\end{equation}
where $t_{\rm cycle}$ is the time for one complete cycle, i.e.
from crossing the shock from upstream to downstream, diffusing
back towards the shock and crossing from downstream to upstream,
and finally returning to the shock, and $\langle \Delta E
\rangle$ is the average energy gain per cycle.

By considering the collisionless scattering of relativistic
particles by magnetic scattering centres in the upstream and downstream
plasmas, one finds
\begin{equation}
{\langle \Delta E \rangle \over E}  \simeq {4 (R-1)\over 3 R} {V_{S} \over c}
\end{equation}
where $V_s$ is the shock velocity, $R=u_1/u_2$ is the compression
ratio, $u_1=V_s$ and $u_2$ are the upstream and downstream
velocities in the rest frame of the shock.  For a strong shock in
a monatomic gas (e.g. fully ionized plasma) $R=4$.  The
acceleration rate at energy $E$ is defined by 
\begin{equation}
r_{\rm acc} = {1 \over E} {dE \over dt}
\end{equation}

The rate of loss of accelerated particles downstream is the probability
of escape per shock crossing divided by the cycle time
\begin{equation}
r_{\rm esc} = {{\rm Prob.(escape)} \over t_{\rm cycle}}.
\end{equation}
By considering the rate at which cosmic rays cross the shock from
upstream to downstream, and rate of convection of cosmic rays
downstream one obtains the probability of escape downstream per shock
crossing
\begin{equation}
{\rm Prob.(escape)}  \simeq {4 \over R} {V_{S} \over c}
\end{equation}

We see immediately that the ratio of the escape rate to the
acceleration rate depends on the compression ratio
\begin{equation}
{r_{\rm esc} \over r_{\rm acc}} \approx {3 \over R-1}
\end{equation}
and for a strong shock ($R=4$) the two rates are equal.  As we
shall see below, a consequence of this is that the asymptotic
spectrum of particles accelerated by a strong shock is the well-known
$E^{-2}$ power-law.

The cycle time depends on the diffusion coefficients
upstream, $k_1$, and downstream, $k_2$, of the shock, and on the
shock velocity and compression ratio and is given by
\begin{equation}
t_{\rm cycle} \approx {4 \over c} \left( {k_1 \over u_1} + {k_2
\over u_2} \right).
\label{eq:cycle}
\end{equation}
The acceleration rate is then given by
\begin{equation}
r_{\rm acc} \approx {(R-1)u_1 \over 3R} \left( {k_1 \over u_1} + {k_2
\over u_2} \right)^{-1}.
\end{equation}
Assuming the the diffusion coefficients upstream and downstream
have the same power-law dependence on energy, e.g.,
\begin{equation}
k_1 \propto k_2 \propto E^{\delta},
\end{equation}
then the acceleration rate also has a power-law dependence
\begin{equation}
r_{\rm acc}\propto E^{-\delta}.
\end{equation}
Note that more correctly the diffusion coefficient will be a
function of magnetic rigidity, $\rho$, which, for
ultra-relativistic particles considered in this paper, is
approximately equal to $E/Ze$ where $Ze$ is the charge.  However,
in this paper since we are mainly concerned with singly charged
particles we shall work in terms of $E$ rather than rigidity.

The leaky-box acceleration model may then be considered as
follows.  A particle of energy $E_0$ is injected into the leaky
box.  While inside the box, the particle's energy changes at a
rate $dE/dt = E r_{\rm acc}(E)$ and that in any short time
interval $\Delta t$ the particle has a probability of escaping
from the box given by $\Delta t r_{\rm esc}(E)$.  The energy
spectrum of particles escaping from the box then approximates the
spectrum of shock accelerated particles.

\subsection{Energy spectrum for acceleration rate different to escape rate}

Let us consider first the case of no energy losses, or losses due
to any other process.  $N_0$ particles of energy $E_0$ are
injected at time $t=0$, and we assume the following acceleration
and escape rates:
\begin{equation}
r_{\rm acc} = a E^{-\delta},
\end{equation}
\begin{equation}
r_{\rm esc} = c E^{-\delta}.
\end{equation}
The energy at time $t$ is then obtained simply by integrating
\begin{equation}
dE/dt = aE^{(1-\delta)},
\label{eq:accrate}
\end{equation}
giving
\begin{equation}
E(t) = (E_0^\delta + \delta a t)^{1/\delta}.
\label{eq:E(t)}
\end{equation}
The number of particles remaining inside
the accelerator at time $t$ after injection is obtained by solving
\begin{equation}
dN/dt = - N(t) c [E(t)]^{-\delta}.
\end{equation}
Using Eq.~\ref{eq:E(t)} and integrating, one has  
\begin{equation}
\int_{N_0}^{N(t)} N^{-1} dN = - c \int_0^t (E_0^\delta + \delta a
t)^{-1} dt,
\end{equation}
giving
\begin{equation}
N(t) = N_0 [E(t) / E_0]^{-c/a}.
\end{equation}
Since $N_0 - N(t)$ particles have escaped from the accelerator
before time $t$, and therefore have energies between $E_0$ and
$E(t)$, the differential energy spectrum of particles which
have escaped from the accelerator is simply given by
\begin{eqnarray}
dN/dE & = & N_0 (\Gamma - 1) (E_0)^{- 1} (E/E_0)^{- \Gamma}, 
\hspace{10mm} (E>E_0)
\label{eq:spec_nocut}
\end{eqnarray}
where $\Gamma = (1+c/a)$ is the differential spectral index.  We
note that for $r_{\rm esc}(E) = r_{\rm acc}(E)$ one obtains the
standard result for acceleration at strong shocks $\Gamma = 2$.

\section{Cut-off due to finite acceleration volume, etc.}

Even in the absence of energy losses, acceleration usually ceases
at some energy due to the finite size of the acceleration volume
(e.g. when the gyroradius becomes comparable to the characteristic
size of the shock), or as a result of some other process.  We
approximate the effect of this by introducing a constant term to
the expression for the escape rate:
\begin{equation}
r_{\rm esc} = c E^{-\delta} + c E_{\rm max}^{-\delta}.
\label{eq:escrate}
\end{equation}
where $E_{\rm max}$ is defined by the above equation and will
be close to the energy at which the spectrum steepens due
to the constant escape term.  We shall refer to $E_{\rm max}$ 
as the ``maximum energy'' even though some particles will be
accelerated to energies above this.

Following the same procedure as for the case of a purely
power-law dependence of the escape rate, we obtain the
differential energy spectrum of particles escaping from the
accelerator,
\begin{eqnarray}
{dN \over dE} &=& N_0 (\Gamma - 1) (E_0)^{- 1} (E/E_0)^{- \Gamma}
\nonumber \\ && \times [ 1 + ( {E / E_{\rm max}} )^\delta ] \exp
\left\{ - {\Gamma - 1 \over \delta} \left[ \left( {E \over E_{\rm
max}} \right)^\delta - \left( {E_0 \over E_{\rm max}}
\right)^\delta \right] \right\}.
\hspace{5mm} (E>E_0)
\end{eqnarray}
for $\delta>0$.  For $\delta = 0$ we note that from
Eq.~\ref{eq:escrate} $r_{\rm esc}=(2c)$ at all energies.
Thus the situation is equivalent to the case of $E_{\rm max} \to
\infty$ and the spectrum is given by Eq.~\ref{eq:spec_nocut}
provided we replace $c$ with $2c$, i.e. we must replace $\Gamma$
with $\Gamma' =(2 \Gamma-1)$.  

\begin{figure}[htb]
\vspace{17.0cm}
\includegraphics{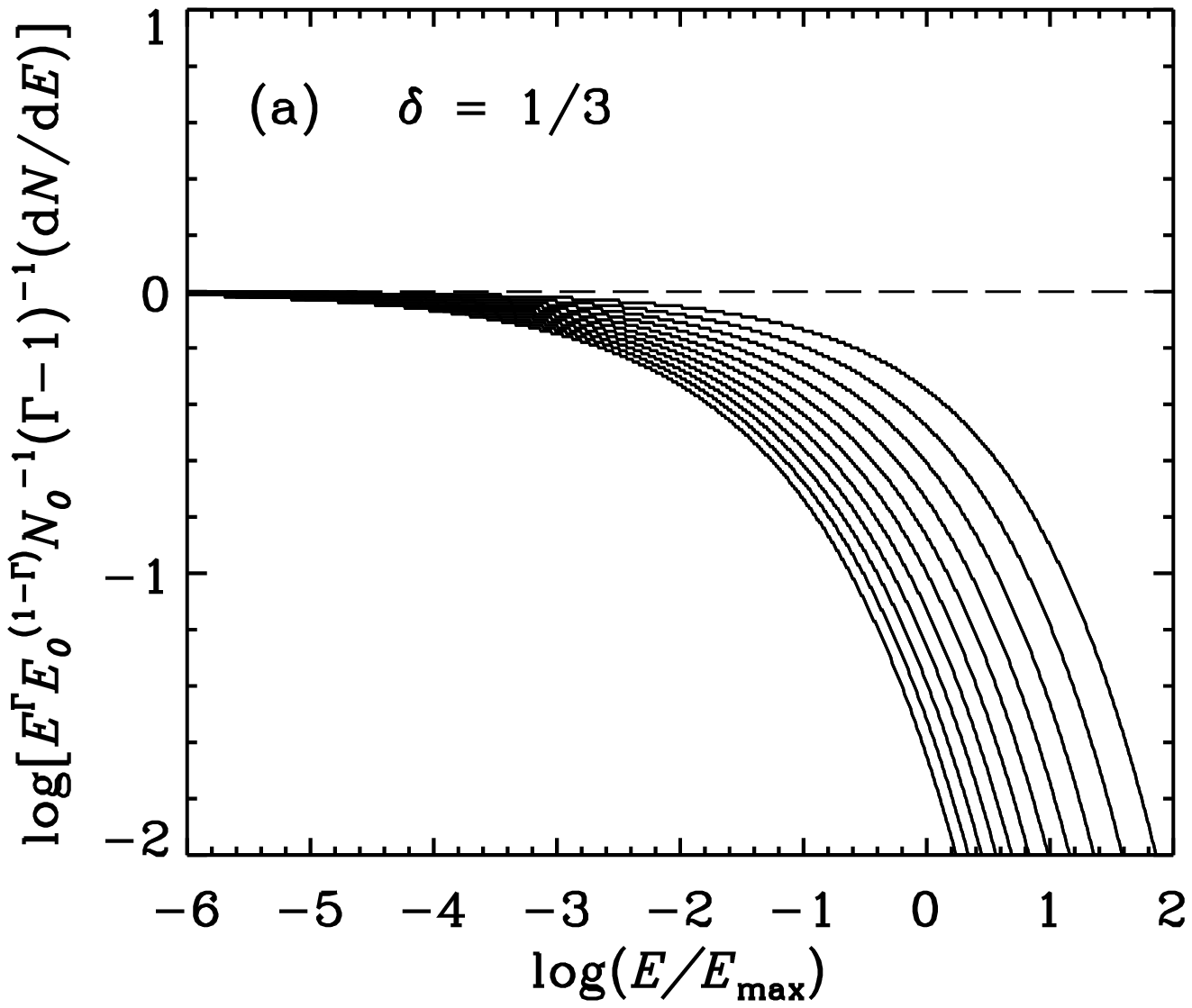}
\includegraphics{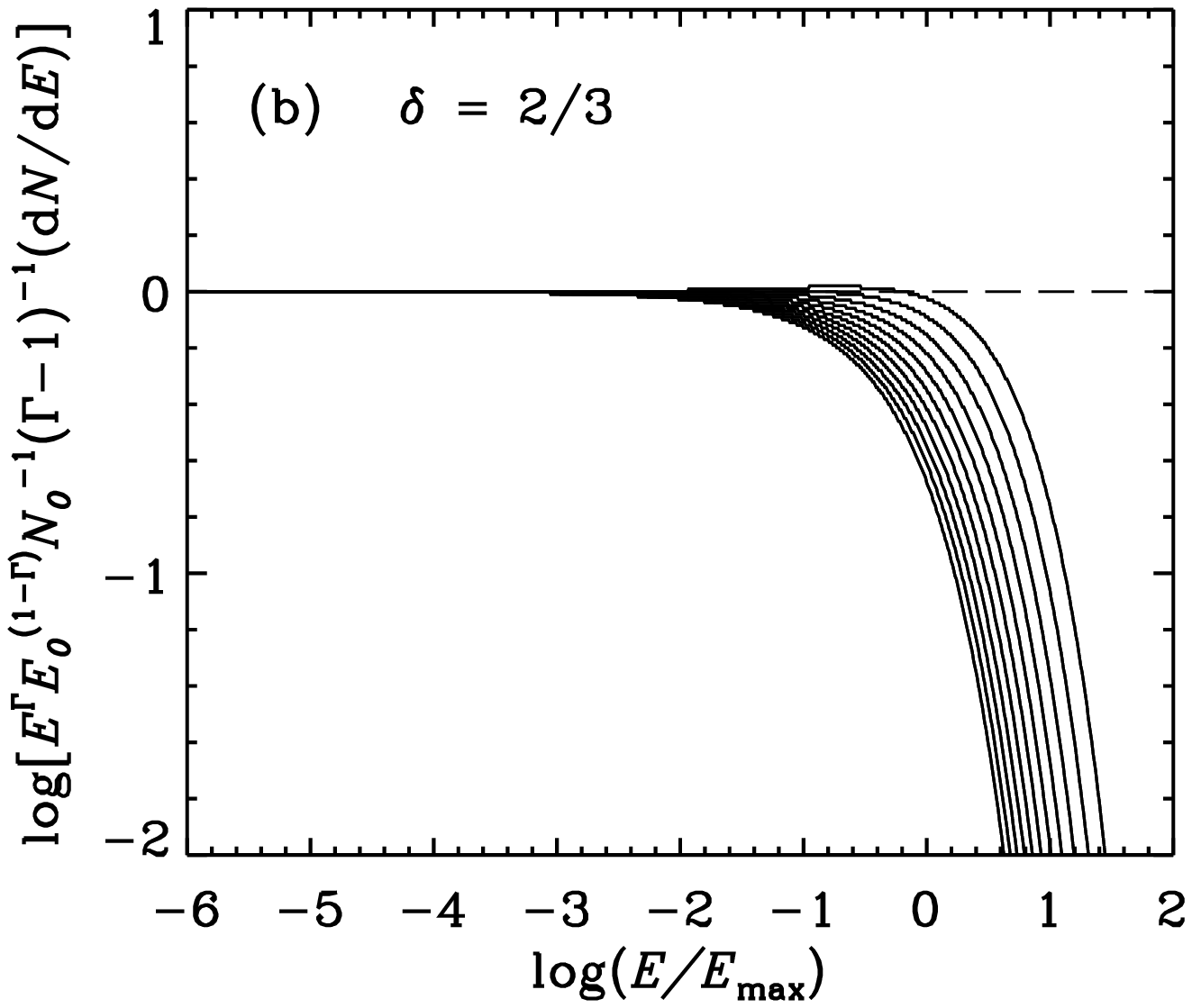}
\includegraphics{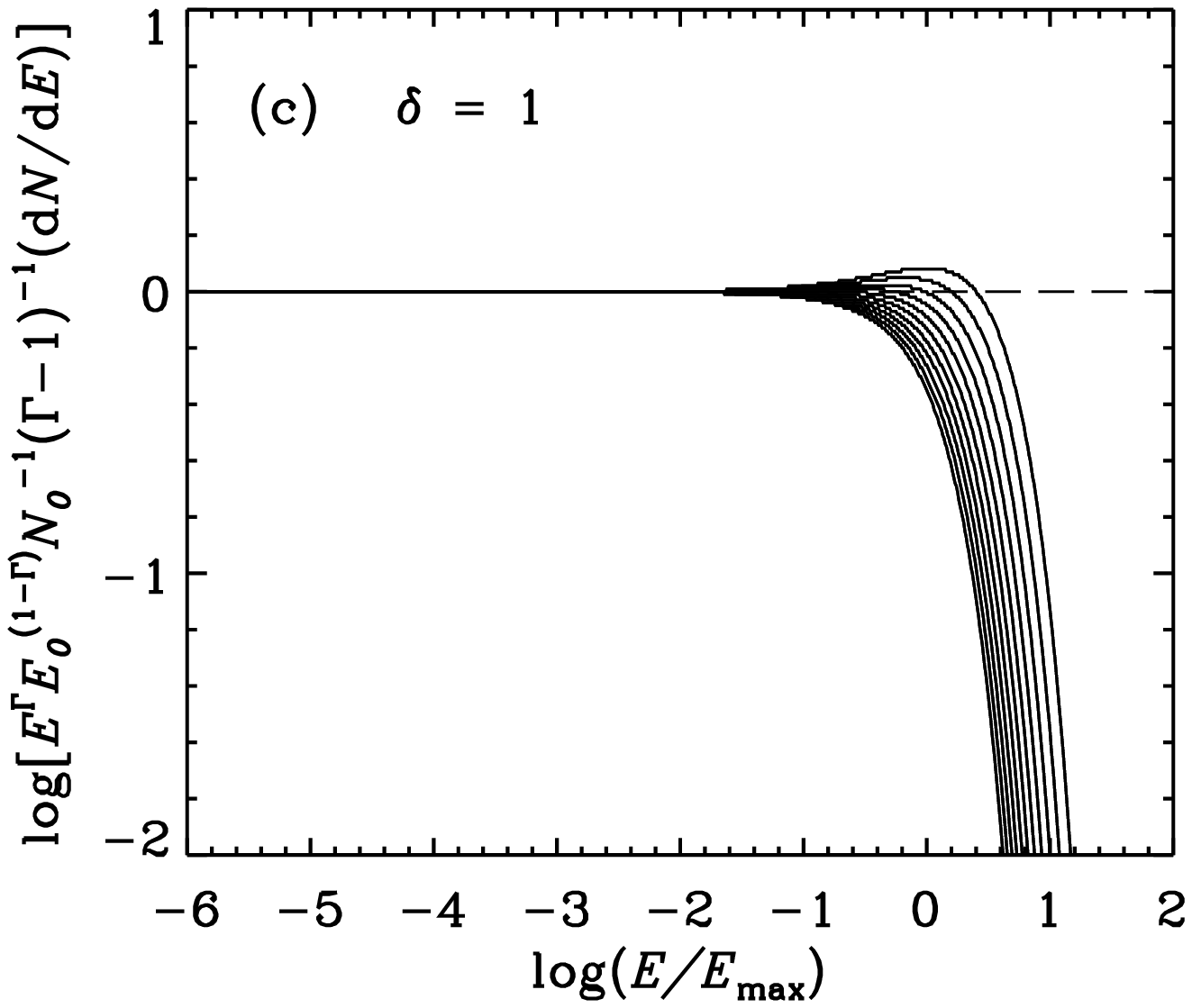}
\caption{Differential energy spectrum for (a) $\delta=1/3$, (b)
$\delta=2/3$, (c) $\delta=1$.  In each case we show results for
$\Gamma=1.5$ (highest curve), 1.6, 1.7, \dots, 2.5 (lowest
curve).}
\label{leakage_cutoff}
\end{figure}

\begin{figure}[htb]
\vspace{12.0cm}
\includegraphics{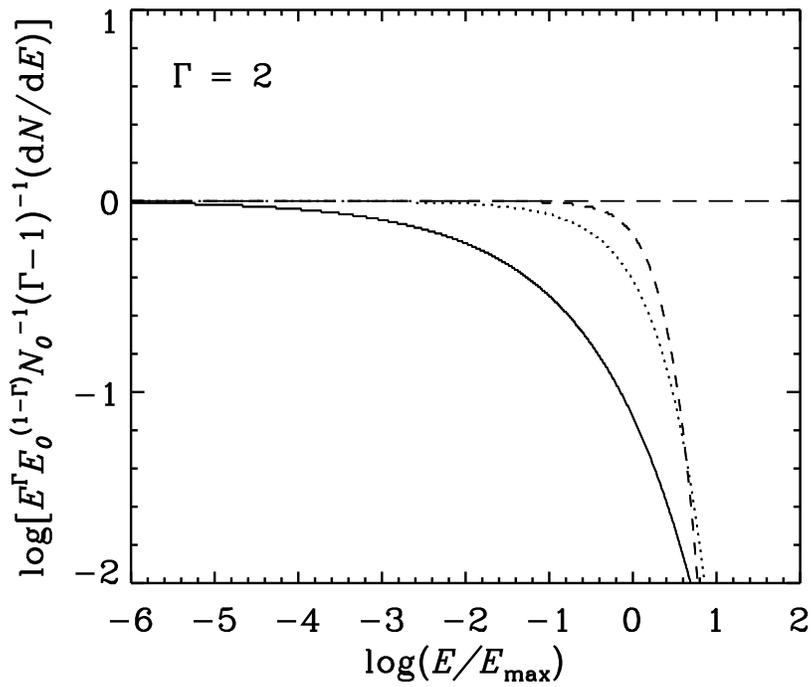}
\caption{Differential energy spectrum for $\Gamma=2$ and
$\delta=1/3$ (solid curve), 2/3 (dotted curve) and 1 (dashed
curve).}
\label{leakage_cutoff_comp}
\end{figure}

We show in Fig~\ref{leakage_cutoff} the differential energy
spectrum for $\Gamma$ ranging from 1.5 to 2.5, for (a)
$\delta=1/3$, (b) $\delta=2/3$, (c) $\delta=1$.  In
Fig~\ref{leakage_cutoff_comp} we compare the spectra for $\Gamma=2$
and $\delta$ ranging from $1/3$ to $1$, and note that the energy
dependence of the diffusion coefficient has a profound influence
on the shape of the cut-off.

Spectra such as those presented in Fig.~\ref{leakage_cutoff} 
may be used to model the source spectra of high energy cosmic ray nuclei
of various species if one replaces energy with magnetic
rigidity, $\rho = pc/Ze \approx E/Ze$, where $Z$ is the atomic number:
\begin{eqnarray}
{dN \over d\rho} &=& N_0 (\Gamma - 1) (\rho_0)^{- 1}
(\rho/\rho_0)^{- \Gamma} \nonumber \\ && \times [ 1 + ( {\rho /
\rho_{\rm max}} )^\delta ] \exp \left\{ - {\Gamma - 1 \over
\delta} \left[ \left( {\rho \over \rho_{\rm max}} \right)^\delta
- \left( {\rho_0 \over \rho_{\rm max}} \right)^\delta \right]
\right\}.
\hspace{5mm} (\rho>\rho_0)
\end{eqnarray}

\section{Cut-off due to $E^2$ energy losses}

$E^2$ energy losses of electrons result from synchrotron
radiation and inverse Compton scattering of low-energy photons in
the Thomson regime.  This may be treated as a continuous energy
loss process.  At higher energies, when the electron rest frame
energy of the emitted photon is no longer much less than the
electron rest mass energy (Klein-Nishina regime of Compton
scattering, or quantum synchrotron regime), the energy losses are
no longer continuous and this will be discussed in Section~5.

The introduction of continuous energy losses into the problem is,
in principle, straightforward and accomplished simply by modifying
Eq.~\ref{eq:accrate},
\begin{equation}
dE/dt = a E^{(1-\delta)} -b E^2.
\label{eq:acclossrate}
\end{equation}
For this case, however, the problem is easier to solve using the
Green's function approach adopted by Stecker~\cite{Stecker71}
when considering the ambient spectrum of cosmic ray electrons and
galactic gamma-rays.  Using the appropriate Green's function one
can obtain the steady-state spectrum of particles inside the
leaky-box accelerator, and multiplying this by the escape rate
one obtains the spectrum of particles leaving the accelerator,
\begin{eqnarray}
{d N \over dE} = {\frac{c\, {E^{-\delta}} + {E_{\rm max}^{-\delta}} }
       {a\,{E^{1 - \delta}}- b\,{E^2}}} \exp [ - I(E)]
\end{eqnarray}
where
\begin{eqnarray}
I(E) & \equiv &
\int_{E_0}^E {\frac{c\, {E^{-\delta}} + {E_{\rm max}^{-\delta}} }
       {a\,{E^{1 - \delta}}- b\,{E^2}}}\,dE
\end{eqnarray}
is given in Appendix~A.  The result depends on the parameters
$\delta$, $\Gamma$, $E_0$, $E_{\rm cut}$, $E_{\rm max}$.

\begin{figure}[htb]
\vspace{12.0cm}
\includegraphics{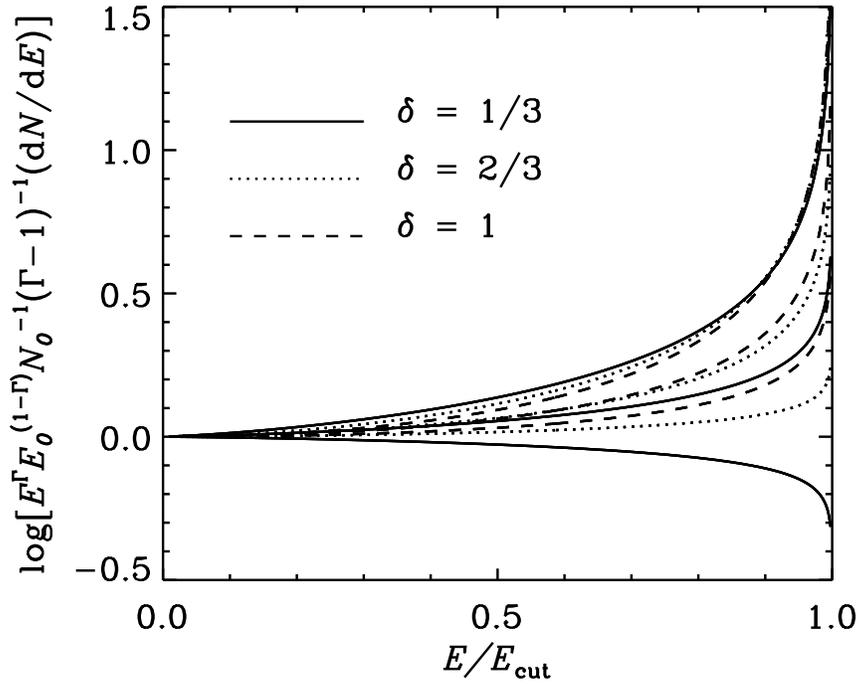}
\caption{Differential energy spectra for $E_0 \ll E_{\rm cut} \ll
E_{\rm max}$ for $\delta=1/3$ (solid curves), 2/3 (dotted curves)
and 1 (dashed curves), and $\Gamma=1.5$ (upper curves), 2.0
(middle curves) and 2.5 (lower curves).}
\label{synch_pileup}
\end{figure}

As a result of the energy loss by particles near the maximum
energy a pile-up in the spectrum may be produced just below
$E_{\rm cut}$.  The size of the pile-up will be determined by the
relative importance of $r_{\rm acc}$ and $r_{\rm esc}$ at
energies just below $E_{\rm cut}$, and so should depend only on
$\Gamma$ and $\delta$ provided $E_0 \ll E_{\rm cut} \ll E_{\rm
max}$.  In this case, the shape of the spectrum is given by
\begin{eqnarray}
{d N \over dE} = N_0 (\Gamma - 1) (E_0)^{- 1} \left( {E \over
E_0} \right)^{- \Gamma} \left[ 1 - \left( {E \over E_{\rm cut}}
\right)^{(1 + \delta)} \right]^{(\Gamma -2 -\delta)/(1 + \delta)}.
\end{eqnarray}
We compare in Fig.~\ref{synch_pileup} the spectra for this case
for $\delta=1/3$, 2/3 and 1, and for $\Gamma = 1.5$, 2, and 2.5,
and note that the pile-ups are higher for flatter spectra, and
that for steep spectra the pile-up may be absent or the spectrum
may steepen before the cut-off if $\delta$ is small.  The effect
of synchrotron losses on multiple diffusive shock acceleration
have been considered by Melrose and Crouch\cite{MelroseCrouch},
and similar effects were found.

\begin{figure}[htb]
\vspace{17.0cm}
\includegraphics{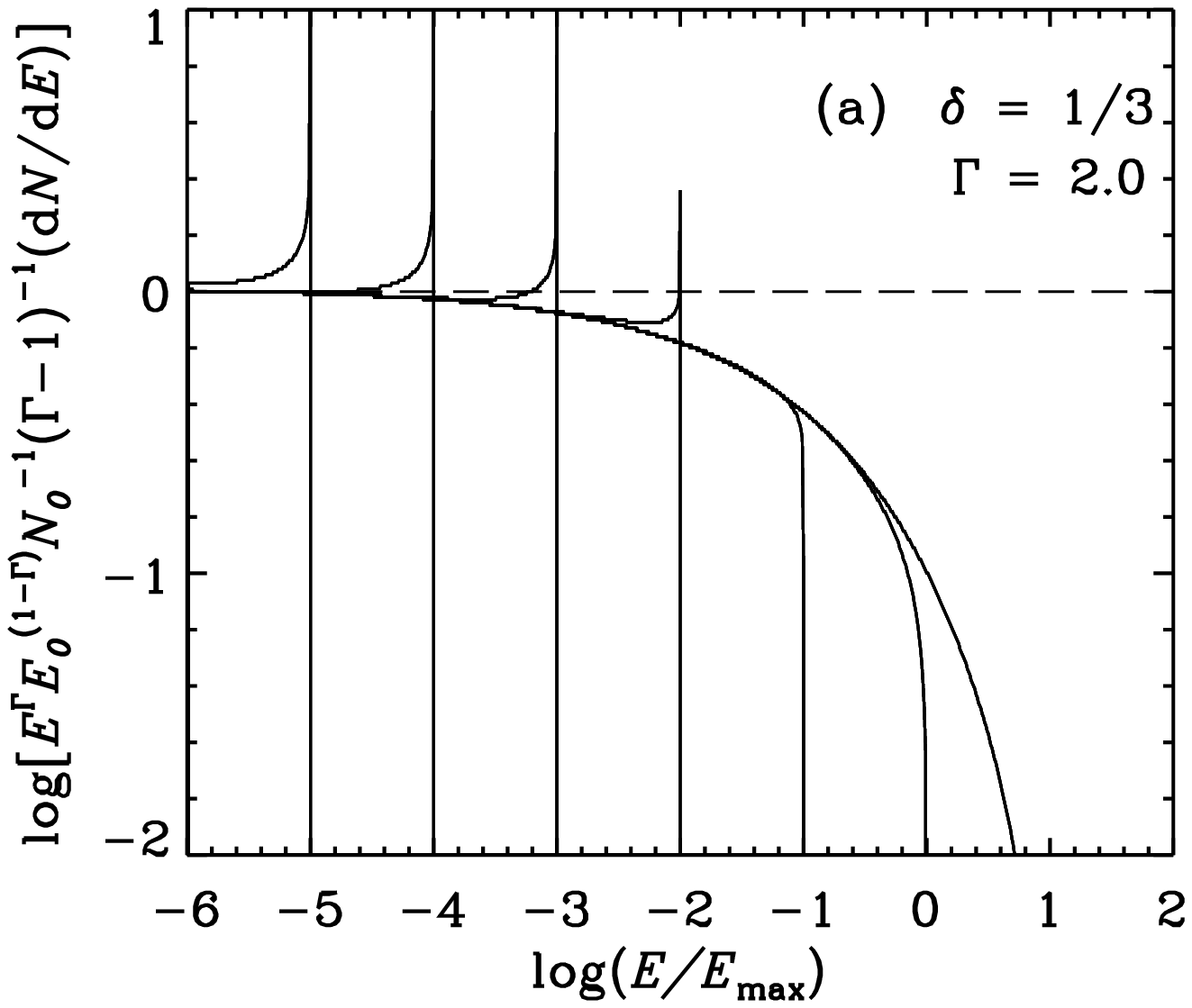}
\includegraphics{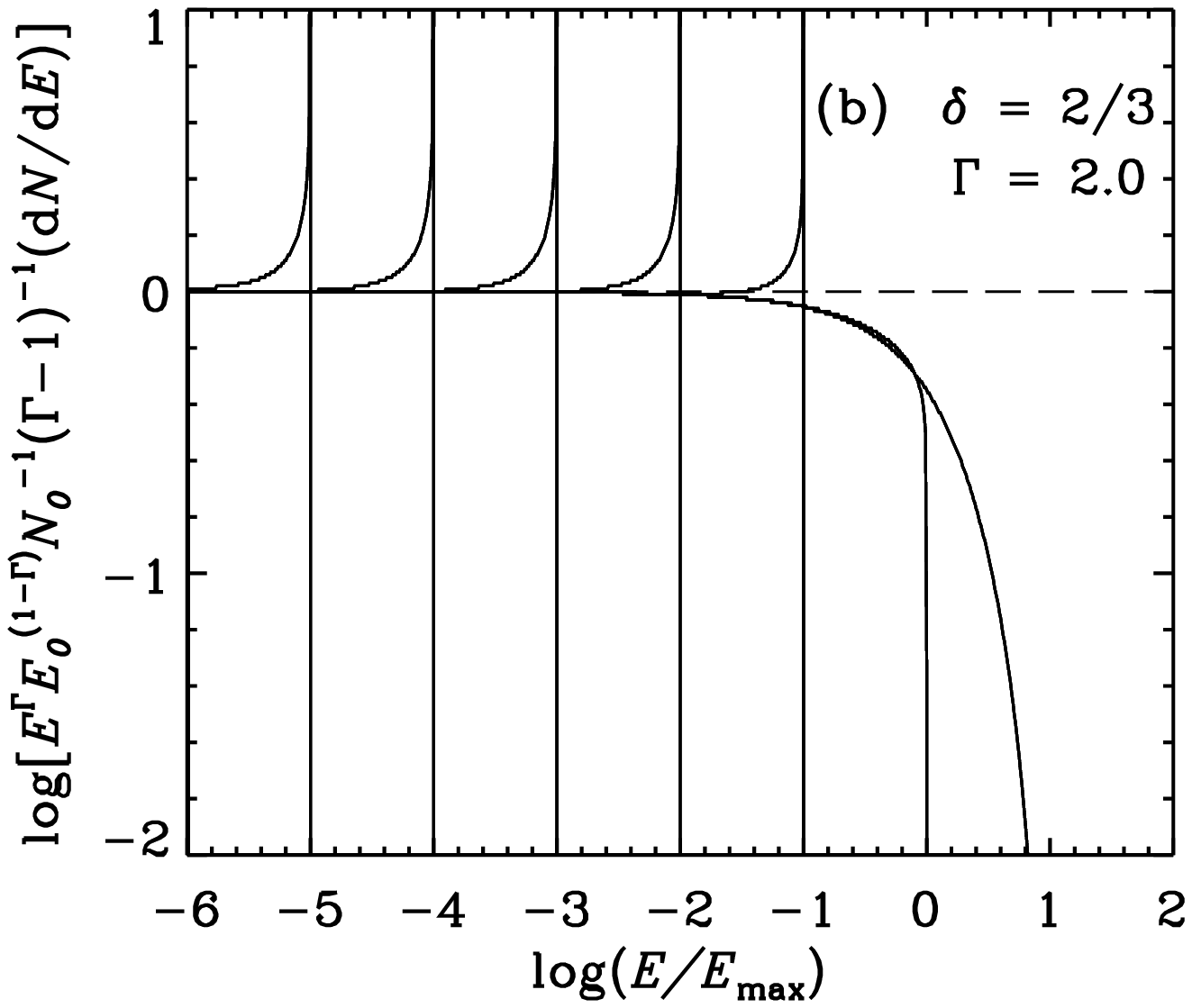}
\includegraphics{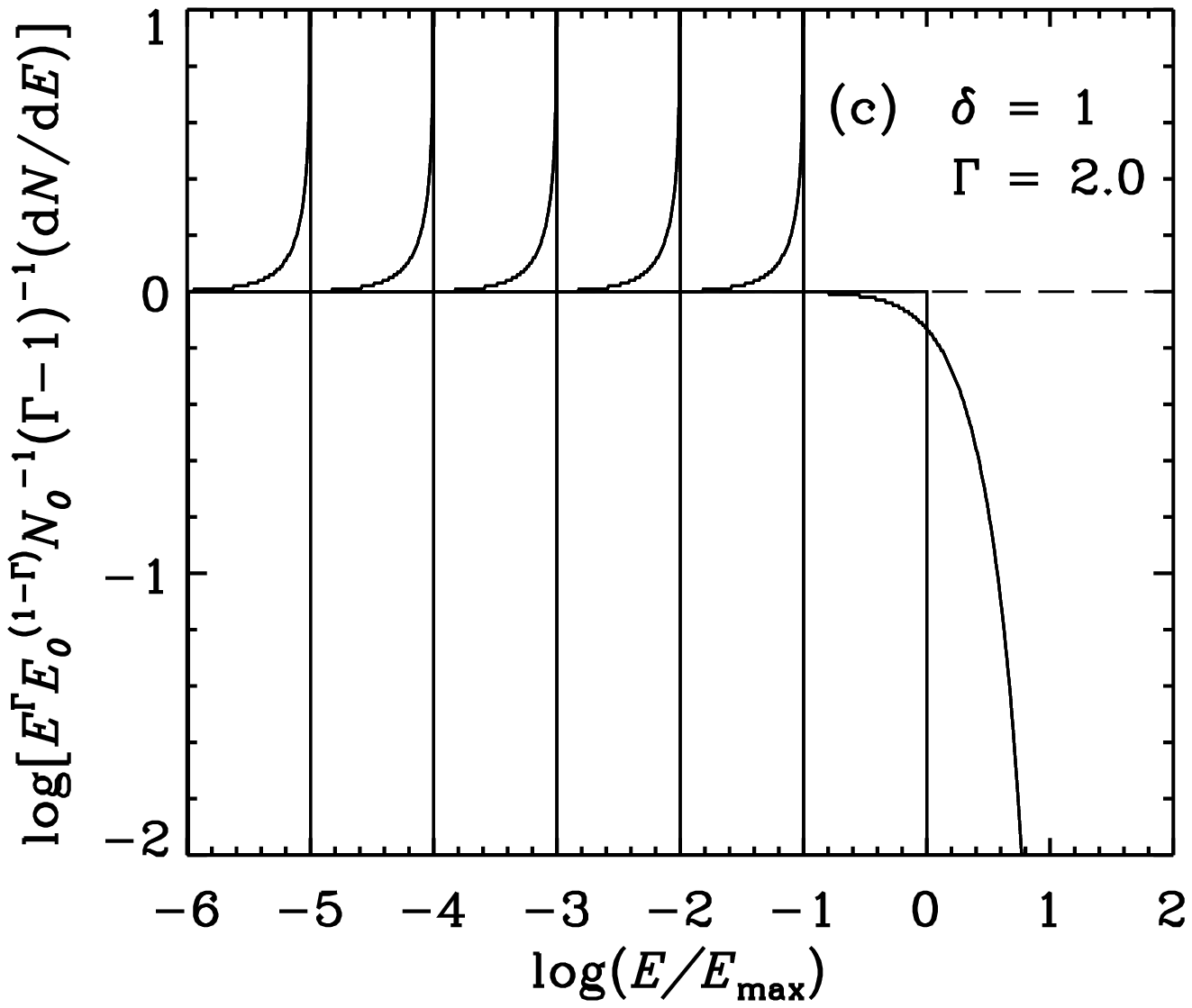}
\includegraphics{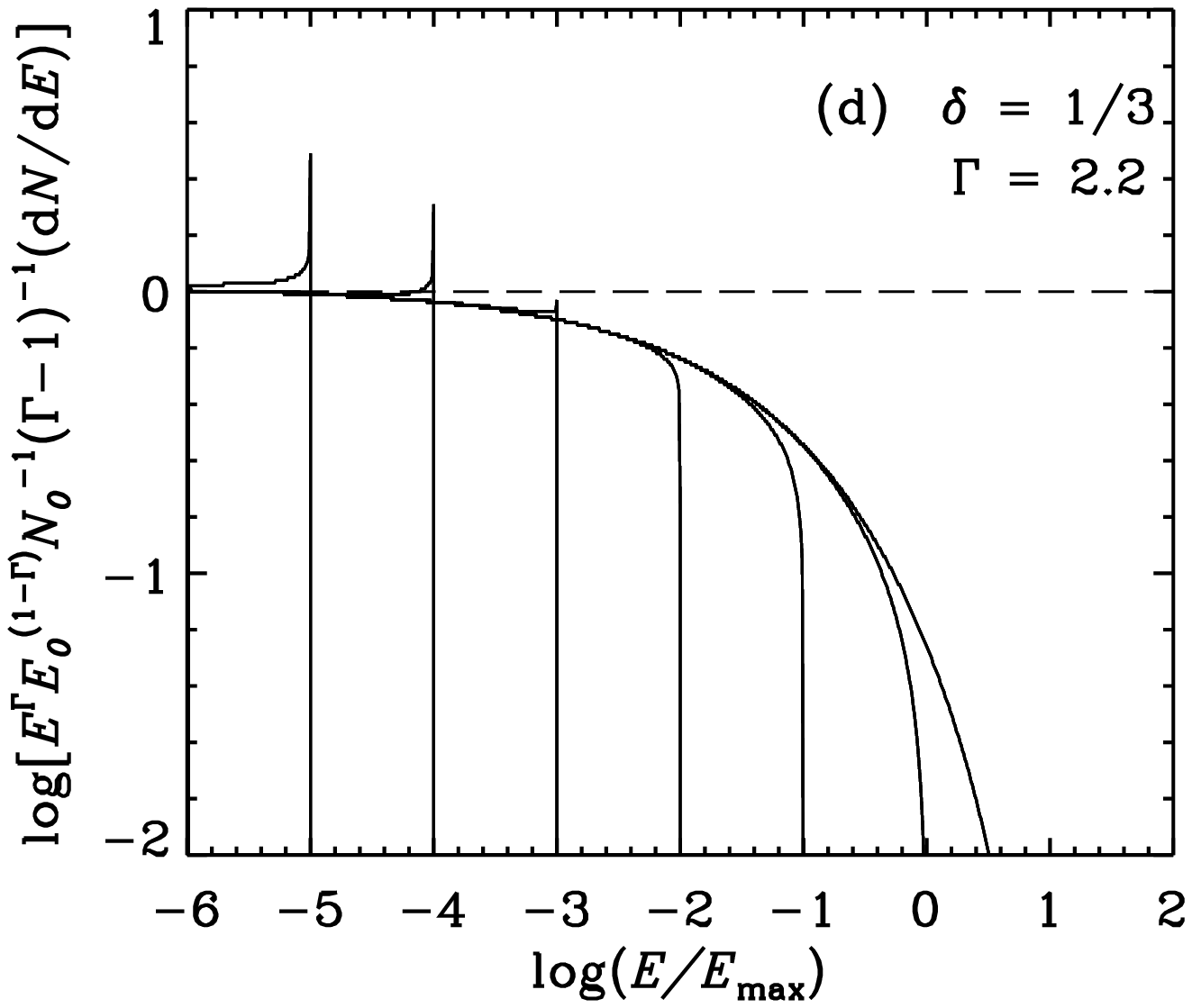}
\caption{Differential energy energy spectrum for $\Gamma=2$ for
(a) $\delta=1/3$, (b) $\delta=2/3$, (c) $\delta=1$, and (d) for
$\Gamma=2.2$ and $\delta=1/3$.  In each case we show results for
$E_{\rm cut}/E_{\rm max} = 10^{-5}$ (leftmost curve), $10^{-4}$,
\dots , $10^{1}$ (rightmost curve).}
\label{synch_cutoff}
\end{figure}

For the case where $E_{\rm cut}$ is comparable to or not much
less than $E_{\rm max}$ we show in Fig~\ref{synch_cutoff} the
differential energy spectrum for $\Gamma=2$ for (a) $\delta=1/3$,
(b) $\delta=2/3$, (c) $\delta=1$, and in each case we show
results for $E_{\rm cut}/E_{\rm max}$ ranging from $10^{-5}$ to
$10^{1}$.  Figs.~\ref{synch_cutoff}(a)---(c) are identical in the
sense that they are all for $\Gamma=2$ and show synchrotron
cutoffs at 1, 1/10, ... 10$^{-5}$ of $E_{\rm max}$. The only
difference is in the energy dependence of the diffusion
coefficient.  In Fig.~\ref{synch_cutoff}(c), for $\delta=1$, the
shape of the pile-up is independent of $E_{\rm cut}$ for $E_{\rm
cut} \le 0.1 E_{\rm max}$.  In contrast, in
Fig.~\ref{synch_cutoff}(a), for $\delta=1/3$, the size of the
pile-up decreases with increasing $E_{\rm cut} / E_{\rm max}$,
and so does not depend simply on $\Gamma$ and $\delta$.  This is
because the constant escape term gives rise to a steepening in
the spectrum, and thus from Fig.~\ref{synch_pileup} we expect a
smaller synchrotron pile-up, or even a steepening.  For example,
note that in Fig.~\ref{synch_cutoff}(a), for $\delta=1/3$ the the
spectrum in the absence of synchrotron losses steepens at
significantly lower energies (due to the constant leakage term)
than for the case of $\delta=1$, and at $E = 10^{-2} E_{\rm max}$
has a differential index of $\sim 2.2$.  We plot in
Fig.~\ref{synch_cutoff}(d) results generated for $\Gamma=2.2$ and
$\delta=1/3$ and note that the size of the pile-up for $E_{\rm
cut} \ll E_{\rm max}$ is similar to the case for $\Gamma=2$,
$\delta=1/3$, and $E = 10^{-2} E_{\rm max}$.

\section{Cut-off due to inverse Compton losses on the 2.7~K background}

In this section, we discuss the acceleration of electrons in the
presence of a blackbody radiation field.  Such a scenario is
important in many areas of high energy astrophysics.  For
example, one may have acceleration of electrons in active
galactic nuclei being responsible for radio, optical, X-ray, and
gamma-ray emission, and this acceleration takes place in the
presence of an intense radiation field, e.g. from the accretion
disk.  

We choose as our example, the case where the cosmic microwave
background radiation is the dominant radiation field.  Such a
situation would occur during 1st order Fermi acceleration of
cosmic ray electrons in our Galaxy at a shock in the interstellar
medium due to a young supernova remnant.

\subsection{Times scales for losses and acceleration}

\begin{figure}[htb]
\vspace{17.0cm}
\includegraphics{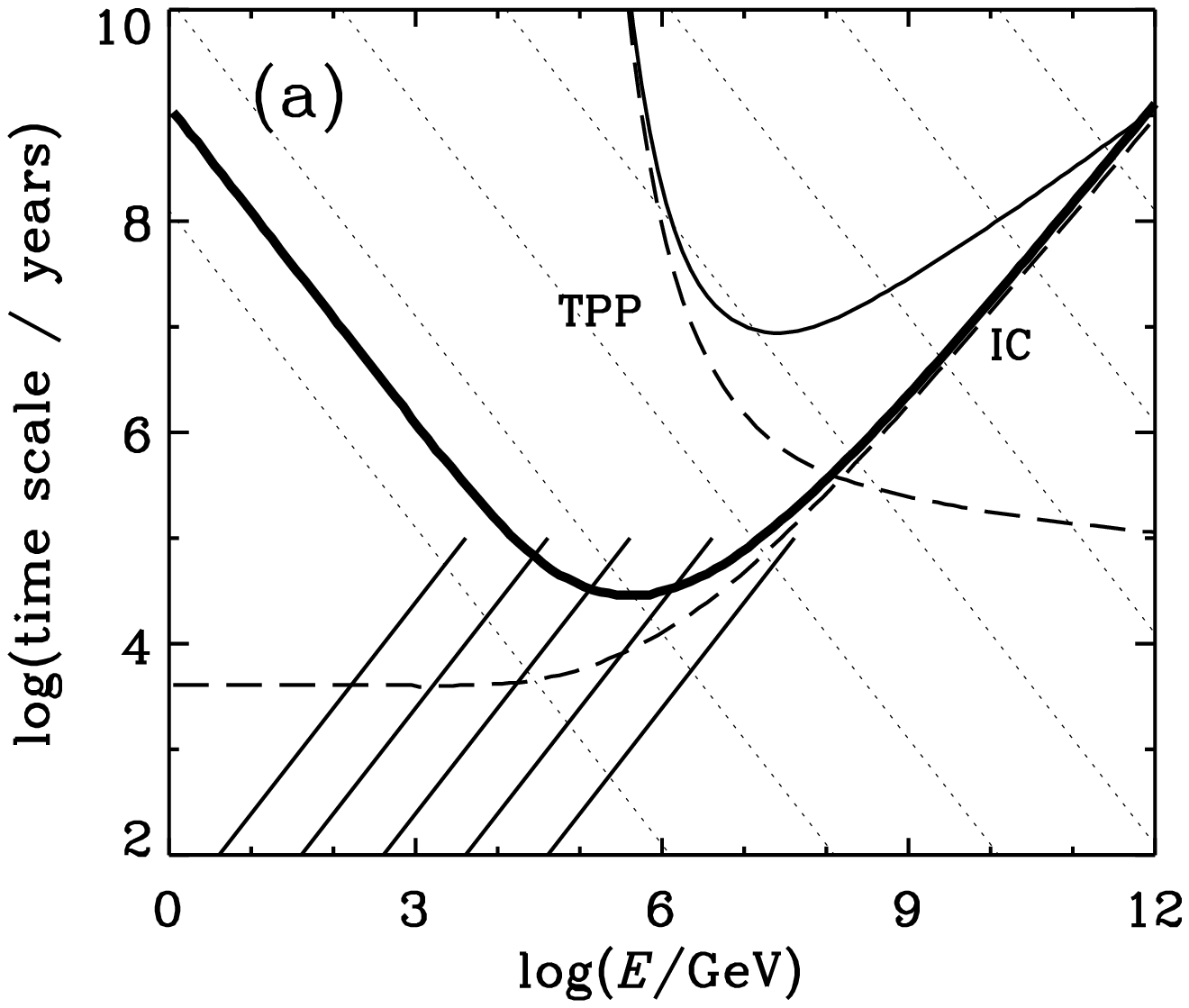}
\includegraphics{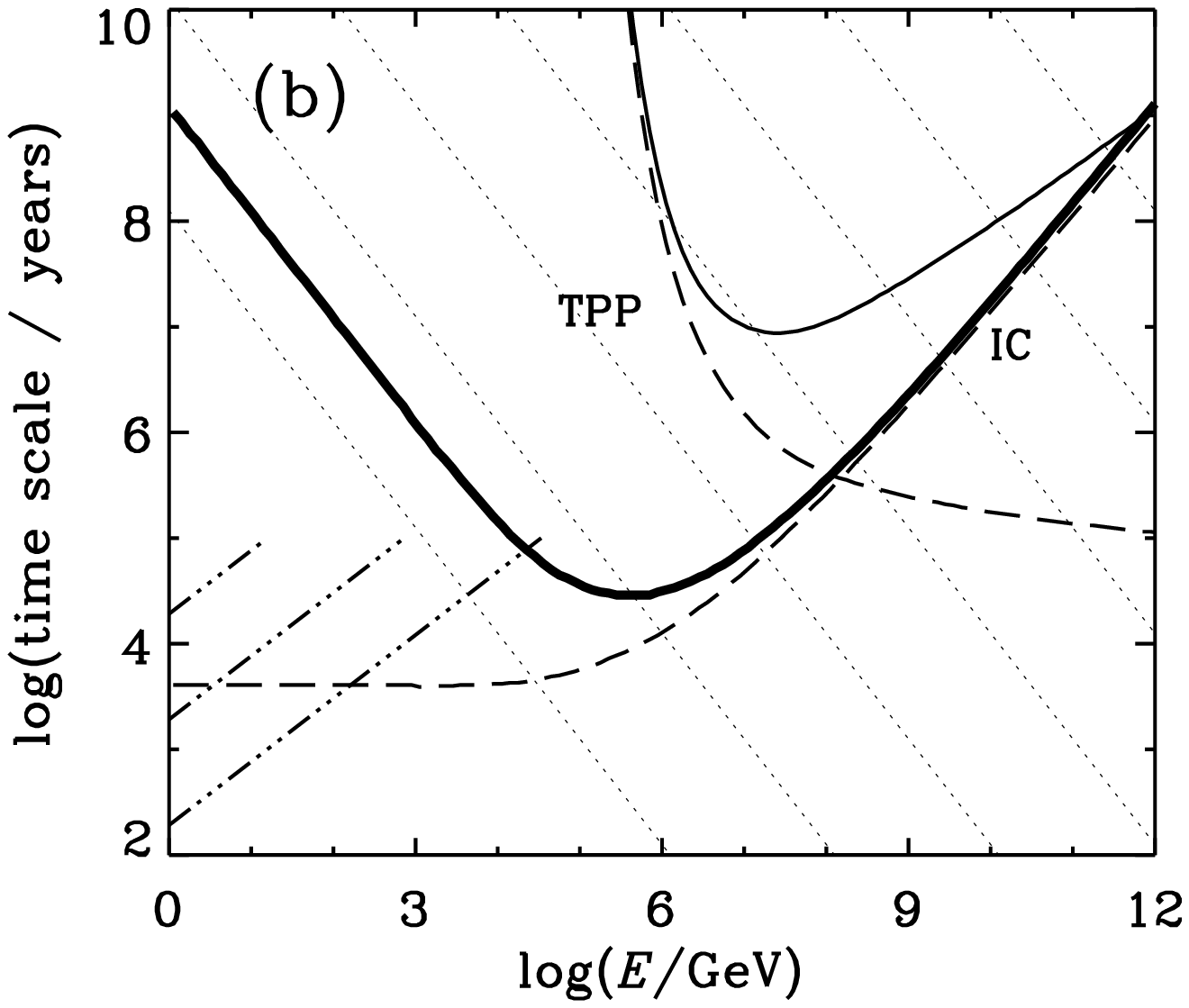}
\includegraphics{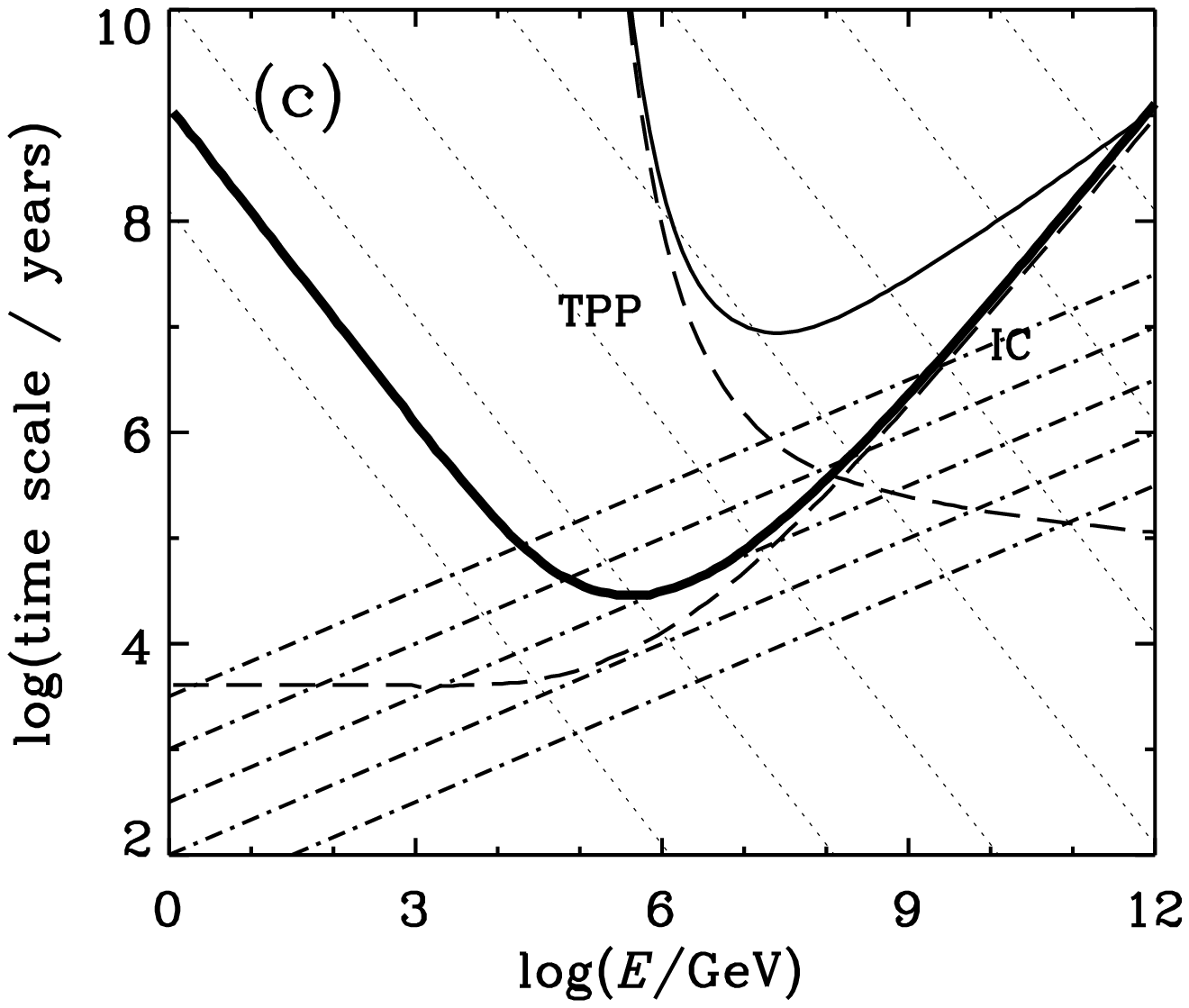}
\caption{The energy loss time scale for synchrotron losses are
shown by thin dotted lines for $B=10 \; \mu$G (lowest line), $1
\; \mu$G, $0.1 \; \mu$G, ... $10^{-11}$~G (highest line). For IC,
the energy loss time scale (thick solid line labeled IC) and mean
interaction time (dashed line labeled IC) are shown; the
corresponding curves labelled TPP are for triplet pair
production.  Acceleration time scales are shown for: (a)
$\delta=1$ (Bohm diffusion) with $\xi B=1.4 \times 10^{-9} \;
\mu$G (leftmost solid line), $1.4 \times 10^{-8}$, ...  $1.4
\times 10^{-5} \; \mu$G (rightmost solid line); (b) $\delta=0.6$
which has been suggested for the propagation of Galactic cosmic
rays and $u_1=0.1c$ (lowest chain line), $0.03c$, and $0.01c$
(highest chain line); (c) $\delta=1/3$ which would arise in the
case of a Kolmogorov spectrum of turbulence -- a range of
acceleration times (chain lines) are shown for the purpose of
illustration.}
\label{accel_bohm}
\end{figure}

The energy loss rate is given by
\begin{eqnarray}
{dE \over dt} = - {4 \over 3} \sigma_T c U \gamma^2
\end{eqnarray}
where $\sigma_T$ is the Thomson cross section, $\gamma$ is the
electron's Lorentz factor, and $U=B^2/8\pi$ for synchrotron
losses and $U$ is the energy density of the radiation field for
inverse Compton losses in the Thomson regime ($U_{\rm rad}=aT^4$
for blackbody radiation).  Thus, at low energies, the time scale
for energy loss of an electron by synchrotron losses is given by
\begin{eqnarray}
{E \over -dE/dt} = 1.3 \times 10^{10} 
\left( {E \over {\rm 1 \; GeV}} \right)^{-1} 
\left( {B \over 1 \; \mu {\rm G}} \right)^{-2}  
\;\;\; {\rm years}, 
\end{eqnarray}
while that for IC scattering is given by
\begin{eqnarray}
{E \over -dE/dt} =1.3 \times 10^{9} 
\left( {E \over {\rm 1 \; GeV}} \right)^{-1} 
\left( {T \over 2.735} \right)^{-4}  
\;\;\; {\rm years}. 
\end{eqnarray}
The energy loss time scale for synchrotron losses is shown in
Figs.~\ref{accel_bohm}(a)---(c) for $B$ ranging from $10^{-11}$~G
to $10^{-5}$~G by the dotted lines.

 For electrons with energies above $\sim 3 \times 10^4$ GeV
 energy losses during inverse Compton scattering on the microwave
 background photons are no longer continuous.  The Klein-Nishina
 cross section must be used, and in each interaction the electron
 loses a substantial fraction of its energy.  Monte Carlo
 simulations using the differential Klein-Nishina cross section
 have been performed and the mean energy loss per interaction
 $\langle \Delta E \rangle$ has been calculated as a function of
 the electron energy.  One can thus obtain an effective
 energy-loss time scale, $E t_{\rm int}/ \langle \Delta E
 \rangle$, where $t_{\rm int}$ is the mean interaction time. See
 \cite{Pro86,Pro90,Pro92,Pro93} for discussion of Monte Carlo
 calculations of inverse Compton scattering.  The energy loss
 time scale and the interaction time for IC losses are shown in
 Figs.~\ref{accel_bohm}(a)---(c) by the thick solid curves, and
 dashed curves labelled `IC' (the thin solid curves, and dashed
 curves labelled `TPP' give the energy loss time scale and the
 interaction time for triplet pair production; see
 e.g. \cite{Mastichiadis94}).

During acceleration, the cut-off energy (and the loss process
responsible) will depend on the strength of the magnetic field
and the acceleration rate.  The acceleration rate is inversely
proportional to the diffusion coefficient (see
Eq.~\ref{eq:cycle}), and so the highest acceleration rate is
achieved with the lowest diffusion coefficient.

The diffusion coefficients required $k_1$ and $k_2$ are the
coefficients for diffusion parallel to the shock normal.  The
diffusion coefficient along the magnetic field direction is some
factor $\eta$ times the minimum diffusion coefficient, known as
the Bohm diffusion coefficient,
\begin{equation}
k_\parallel = \eta {1 \over 3} R_g c
\end{equation}
where $R_g$ is the gyroradius, and $\eta > 1$.

Parallel shocks are defined such that the shock normal is
parallel to the magnetic field ($\vec{B} || \vec{u_1}$).  In this
case, making the approximation that $k_1 = k_2 = k_\parallel$,
$R=4$, and $B_1 = B_2$ one obtains
\begin{equation}
r_{\rm acc}^\parallel \approx {3 \beta_1^2 \over 20 \eta} \omega_g
\end{equation}
where $\beta_1 = u_1/c$ and $\omega_g=c/R_g$.  For a shock speed
of $u_1 = 0.1 c$ and $\eta=10$ one obtains an acceleration rate
\begin{equation}
r_{\rm acc}^\parallel  \approx 1.5 \times 10^{-4} \omega_g.
\end{equation}

For the oblique case, the angle between the magnetic field direction 
and the shock normal is different in the upstream and downstream regions,
and the direction of the plasma flow also changes at the shock.
The diffusion coefficient in the direction at angle $\theta$ 
to the magnetic field direction is given by
\begin{equation}
k = k_\parallel \cos^2 \theta + k_\perp \sin^2 \theta
\end{equation}
where $k_\perp$ is the diffusion coefficient perpendicular to the magnetic field.
Jokipii\cite{Jokipii87} shows that 
\begin{equation}
k_\perp \approx {k_\parallel \over 1 + \eta^2}
\end{equation}
provided that $\eta$ is not too large (values in the range up to 10 appear
appropriate).

In the case of acceleration at perpendicular shocks, Jokipii\cite{Jokipii87}
has shown that acceleration can be much faster than for the parallel case.
Using $k_\perp$ and $B_2 \approx 4 B_1$ and one obtains
\begin{equation}
r_{\rm acc}^\perp \approx {3 \beta_1^2 \eta\over 8} \omega_g.
\end{equation}
For a shock speed of $u_1 = 0.1 c$ and $\eta=10$ one obtains an acceleration
rate of 
\begin{equation}
r_{\rm acc}^\perp \approx 0.04 \omega_g.
\end{equation}

Generally, for stochastic particle acceleration by electric
fields induced by motion of magnetic fields the acceleration rate
is given by
\begin{equation}
r_{\rm acc} = \xi(E) \omega_g \approx 0.01 \xi(E) \left( {E \over {\rm 1 \; GeV}} \right)^{-1} 
\left( {B \over 1 \; \mu {\rm G}} \right) 
\;\;\; {\rm s} 
\end{equation}
where $\xi(E) \le 1$ and depends on the acceleration mechanism;
constant values of $\xi$ ranging from $1.5 \times 10^{-4}$ to
0.04 might be achieved by first order Fermi acceleration with Bohm
diffusion ($k \propto E$, i.e. $\delta=1$).

We show in Fig.~\ref{accel_bohm}(a), along with the energy loss
time scales for IC and synchrotron losses, the acceleration time
scale for $\delta=1$ (Bohm diffusion) for ranges of $\xi B$ which
are probably appropriate for supernova shocks.  Comparing the
energy loss time scale for IC and synchrotron loss we see that if
$B > 3 \, \mu$G the cut off will be determined by synchrotron
losses irrespective of the acceleration rate.  Conversely, if $B
< 3 \, \mu$G the cut off will be determined by IC provided the
acceleration rate is low such that the acceleration time curve
intersects the IC loss curve in the Thomson regime.  However, if
the magnetic field is less than $\sim 1 \, \mu$G and $\xi B$ is
in the range $10^{-8}$ to $10^{-6} \; \mu$G, the cut-off in the
spectrum of accelerated electrons will be determined by IC
scattering in the Klein-Nishina regime.  If the magnetic field
were less than $0.1\, \mu$G, the range of $\xi B$ would be
extended up to $10^{-5} \, \mu$G, but for $\xi B > 1.4 \times
10^{-5} \, \mu$G the cut off will be determined by synchrotron
losses.  If the cut off is be determined by IC in the K-N regime,
the shape of the cut-off would be different from that calculated
for the case of $E^2$ energy losses, and one would need to take
account properly of inverse Compton collisions in the
Klein-Nishina regime. The spectrum of synchrotron radiation and
inverse Compton gamma-rays would also depend on the details of
the electron spectrum cut-off and would therefore differ from the
$E^2$ energy loss case.

In Fig.~\ref{accel_bohm}(b) we show acceleration time scales that
would apply if we used the diffusion coefficient inferred from a
comparison of cosmic ray data with propagation model
calculations~\cite{Porter97}, $k = 2.5 \times 10^{28}(E/3\,{\rm
GeV})^{0.6}$ cm$^2$~s$^{-1}$, and shock velocities ranging from
$0.01c$ to $0.1c$.  In this case, the cut-off energy would
probably be determined by IC in the Thomson regime or by
synchrotron losses.

In Fig.~\ref{accel_bohm}(c) we show acceleration time scales
that would arise in the case of a Kolmogorov spectrum of
turbulence.  A range of acceleration times (chain lines) are
shown for the purpose of illustration only.  The diffusion
coefficient can never be lower than the minimum (Bohm) diffusion
coefficient, and so with increasing energy these curves would
violate the Bohm limit.  Thus the curves shown above $\sim 10^4$
GeV are unrealistic for acceleration at supernova remnant shocks
in our Galaxy with typical galactic magnetic fields.  The curves
in Fig.~\ref{accel_bohm}(c) should bend upwards and smoothly
join an appropriate curve from Fig.~\ref{accel_bohm}(a).  However,
with an acceleration rate such as given by the middle curve, the
cut-off energy determined by IC scattering would not be precisely
determined, and could be anywhere between $3 \times 10^5$ and $3
\times 10^7$ GeV, and this would result in an interesting
spectral shape.  While this scenario does not apply at galactic
supernova shocks, it may well apply in other astrophysical
environments where shocks are present and different radiation
fields exist.

\section{Monte Carlo Simulation}

A set of programs was constructed that simulates the particle
 acceleration in the leaky box acceleration model using the Monte
 Carlo method. In the simulation, a particle is injected with energy
 $E_0$ and the acceleration and escape are simulated for a given
 $E_{\rm max}$ value and specific energy dependence of the diffusion
 coefficient. After the injection with energy $E_0$ the development
 of the electron energy spectrum is followed in time steps $\Delta t$.
 At every step the particles gain energy $\Delta t (dE/dt)$. $\Delta t$
 was chosen to be small, 1/100 of the escape time at $E_0$. The
 spectrum of the escaping particles is calculated by weighting
 the particle energy after each step $i$ by $P_{\rm esc}^i \times
 P^{i-1}$, where $P^{i-1}$ is the probability that the particle
 has remained in the accelerator region for $i-1$ time
 steps. $P_{\rm esc}$ is calculated as a function of the particle
 energy for the chosen energy dependence of the diffusion
 coefficient with a constant $P_{\rm esc} (E_{\rm max})$ added to
 simulate a smooth cutoff in the escaping particle spectrum. The
 maximum acceleration time was calculated so that particles could
 have reached energy of 10$E_{\rm max}$ in the absence of the
 constant term in the escape probability. 

\begin{figure}[htb]
\vspace{12.0cm}
\includegraphics{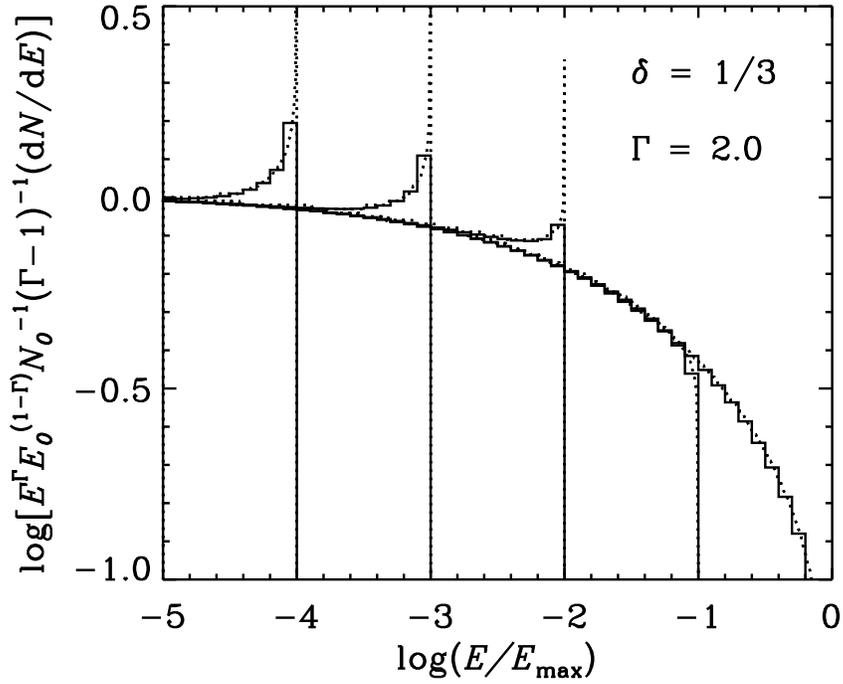}
\caption{Spectrum of accelerated electrons obtained using the 
Monte Carlo technique (histograms) compared with the analytic result
from  Fig.~4(a)
for $E_{\rm max}$ = 10$^6$ GeV and cutoff due to synchrotron energy
loss of 1, 1/10, 1/100, 1/1000, and 1/10000 $E_{\rm max}$. The
diffusion coefficient is proportional to $E^{1/3}$.}
\label{mc_syn_comp}
\end{figure}

 The advantage of the Monte Carlo/numerical approach is that it is
 easily adapted to calculate the escaping electron spectrum for
 both continuous energy losses (e.g. synchrotron or IC in the
 Thomson regime) and non-continuous energy losses (e.g. IC in the
 KN regime).  We next present results obtained by the
 numerical/Monte carlo technique produced in a series of runs made
 including synchrotron energy loss.  The energy loss was
 calculated at every $\Delta t$ and subtracted from the energy
 gain.  The magnetic field value necessary for generating a cutoff
 at 1, 1/10, 1/100, 1/1000, and 1/10000 of $E_{\rm max}$ was
 calculated keeping all `acceleration' parameters as before.  The
 program was tested for the same parameters used in the
 analytical/numerical technique described earlier, and the
 agreement was found to be excellent.  We show, for example, in
Fig.~\ref{mc_syn_comp} the spectrum of accelerated particles in
the presence of synchrotron losses for $\delta=1/3$ obtained
using the Monte Carlo method and compare this with the 
analytic results obtained previously. 

\subsection{Comparison of cutoffs due to synchrotron and IC energy loss}

 Next we used the same procedure to simulate non-continuous
 inverse Compton energy loss. IC in the Klein-Nishina regime was
 incorporated as follows: for every time step, $\Delta t$, the IC
 interaction time was sampled from an exponential distribution
 with mean interaction time, $t_{\rm int}$ as a function of the
 electron energy (given by the long-dashed curve in
 Fig.~\ref{accel_bohm}).  If the sampled interaction time was
 less than $\Delta t$, an IC collision was simulated by the Monte
 Carlo method and the electron energy loss was subtracted from
 the energy gain due to acceleration in the time step. $E_{\rm max}$
 was set to 10$^6$ GeV to enable us to study the non-continuos energy
 loss in the Klein-Nishina regime. All acceleration parameters were
 adjusted to produce an IC energy loss time equal to the acceleration
 time at the same $E_{\rm cut}$ values as in the synchrotron case above.

\begin{figure}
\vspace{12.0cm}
\includegraphics{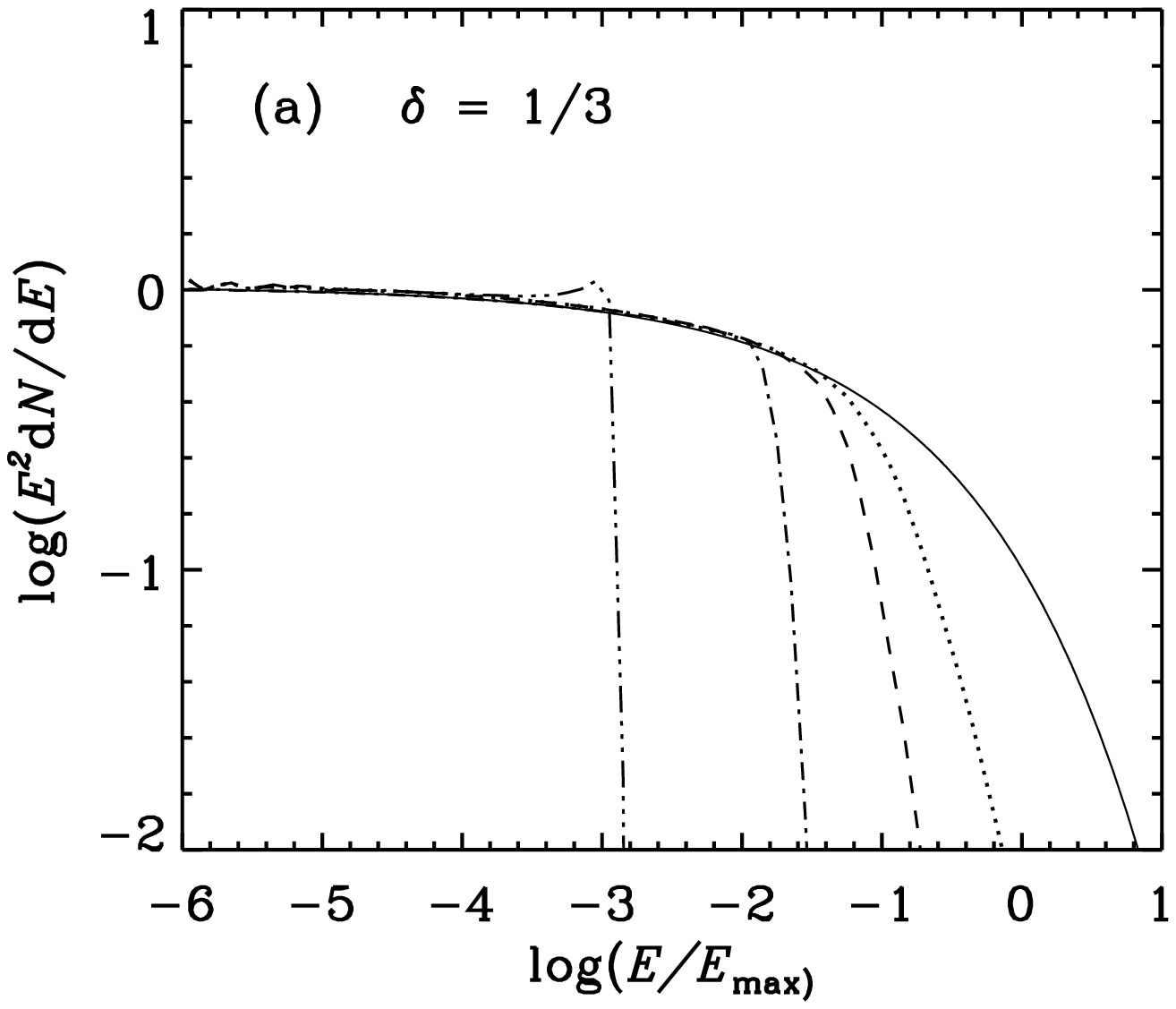}
\includegraphics{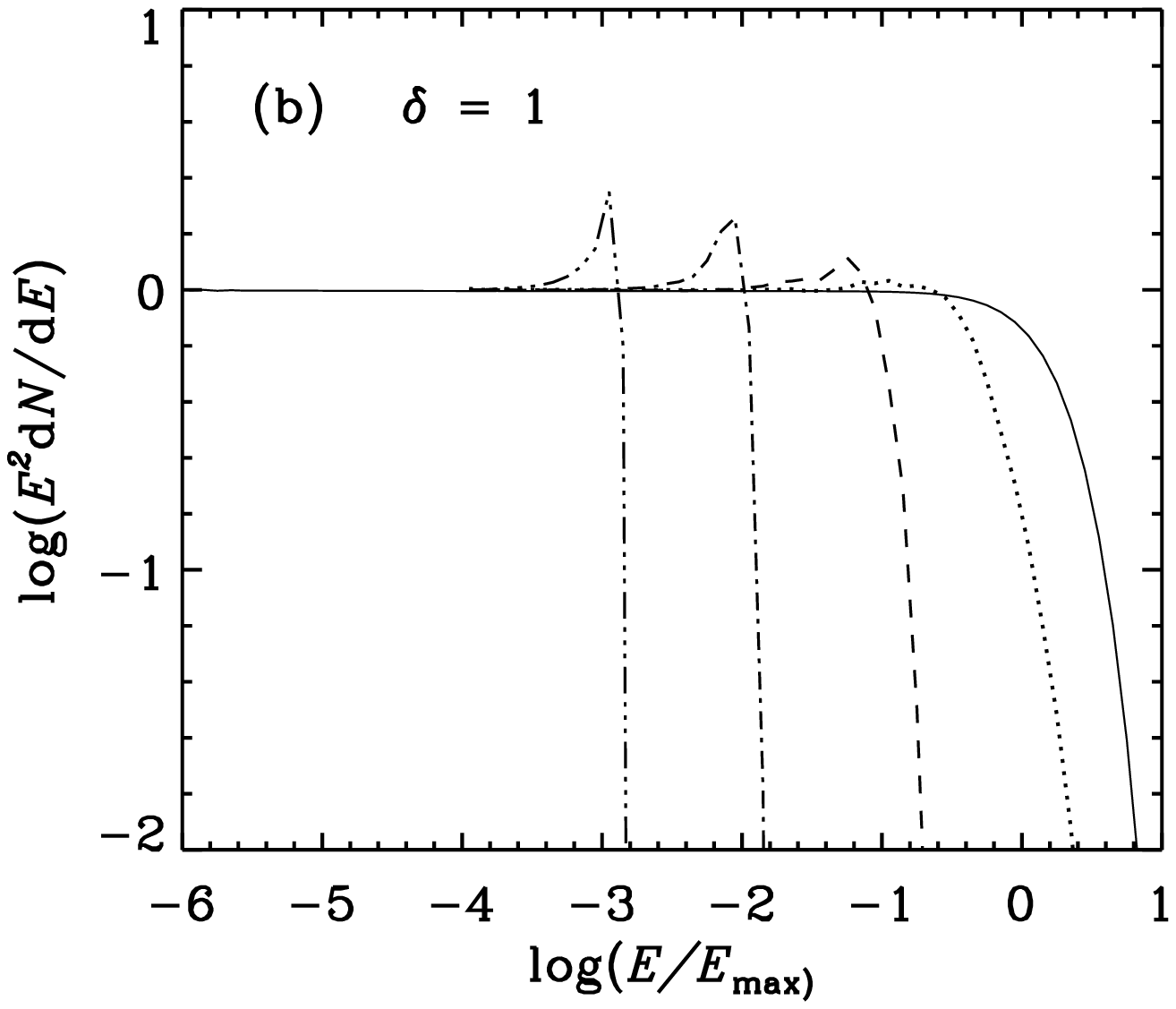}
\caption{Spectrum of accelerated electrons for $E_{\rm max}$ = 10$^6$
GeV and cutoff due to IC energy loss at 1, 1/10, 1/100, and
1/1000 $E_{\rm max}$. The
diffusion coefficient is proportional to (a) $E^{1/3}$, (b) $E$.}
\label{ic_spec}
 \end{figure}

 Figs.~\ref{ic_spec}(a) and \ref{ic_spec}(b) show the electron spectrum in
 the presence of IC energy loss for the case of $\delta=1/3$ and
 $\delta=1$ respectively.  Note that for $E_{\rm cut}$ = 1000
 GeV, where IC is in the Thompson regime and the energy loss is
 quasi-continuous, the picture is similar to that in the
 synchrotron case (Figs.~\ref{synch_cutoff}(a) and (c)). For
 $E_{\rm cut}$ = 10$^6$ GeV, where IC is in the Klein--Nishina
 regime, the shape of the cutoff is qualitatively different and
 much smoother.



 Fig.~\ref{ic_spec}(b) is the version of Fig.~\ref{ic_spec}(a) for the case
 of diffusion coefficient $\propto \, E$. In the case of $E^{1/3}$
 energy dependence of the diffusion coefficient the pile--ups
 are much smaller and almost do not exist for $E_{\rm cut}/E_{\rm max}$
 greater than 10$^{-3}$. Pile--ups are significantly more pronounced
 when the diffusion coefficient is proportional to $E$. The shape
 of these pile--ups, especially for IC losses in the Klein-Nishina
 regime, is very different from  the synchrotron loss case
 (Figs.~\ref{synch_cutoff}(a) and (c)).

\begin{figure}
\vspace{9.3cm}
\includegraphics{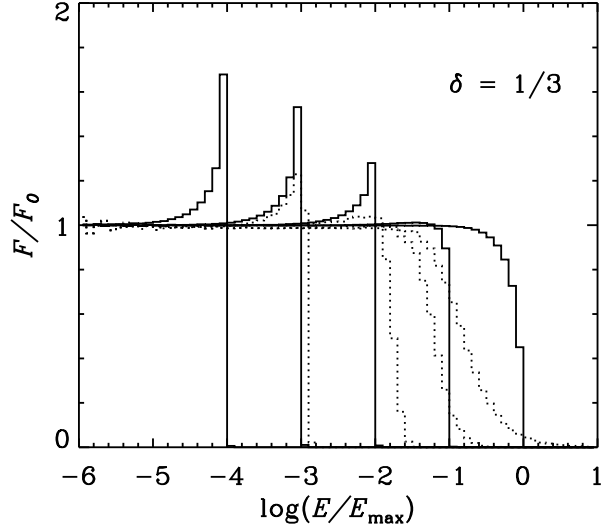}
\caption{Spectrum of accelerated electrons for $E_{\rm max}$ =
 10$^6$ GeV and cutoffs due to synchrotron (solid histogram) and
 IC energy loss (dotted histogram).  Spectrum, divided by that
 for no IC loss, is shown on a linear scale to emphasize the
 cutoffs. The diffusion coefficient is proportional to $E^{1/3}$
 ($F/F_0$ represents the ratio of the spectrum calculated with
 energy losses to the spectrum calculated without energy
 losses).}
\label{ic_syn_comp}
\end{figure}

 Fig.~\ref{ic_syn_comp} compares the cutoffs  due to synchrotron energy
 loss  to those of IC energy loss on a linear scale.
 We show this comparison with a histogram because
 histograms are a better presentation of the Monte Carlo approach
 than smooth curves.

 The dotted histogram in Fig.~\ref{ic_syn_comp} shows the cut--offs
 and pile--ups that are produced by IC energy loss. For
 $E_{\rm cut}/E_{\rm max}$ of 10$^{-3}$ the shape of the pile--up
 in the spectrum of accelerated electrons is not very different
 from the synchrotron loss case. The reason is, of course, that
 the IC energy loss of 1000 GeV electrons in the Thompson regime
 is almost continuous, very similar to the synchrotron loss. 
 For higher ratios, when the IC collisions move deeper and deeper
 in the Klein--Nishina regime and the energy loss becomes a
 significant fraction of the total electron energy, pile--ups
 become very wide, while for $E_{\rm cut}\; = \; E_{\rm max}$ there
 is no pile--up. It is replaced by a spectral cut--off which is very
 different from the cut--offs due to synchrotron loss or to escape
 from the shock.
 
\section{Conclusion}

 We have derived the shape of the spectrum of particles produced
 by 1st order Fermi acceleration near the maximum or cut-off
 energy when the maximum energy is caused by a constant escape
 rate term.  We found that, even in the absence of energy losses,
 the shape of the spectrum of accelerated particles at energies
 well below the nominal maximum energy depends strongly on the
 energy dependence of the diffusion coefficient.  This has
 implications in many areas, for example, in fitting the shape of
 the observed cosmic ray spectrum in the region of the ``knee''
 with model spectra based on power-law acceleration spectra and
 rigidity dependent diffusive escape from the galaxy.

 In the case where the cut-off is due to continuous energy losses
 a prominent pile-up may occur just below the cut-off energy defined
 as the energy at which the total rate of change of energy is
 zero, and the spectrum cuts off sharply at this energy.

 We have developed a Monte Carlo/numerical technique to model
 cut-offs caused by non-continuous energy losses
 (i.e. interactions in which the particle being accelerated loses
 a substantial fraction of its energy). The Monte Carlo technique 
 is tested by comparisons with the analytic calculations in cases
 where possible. Using numerical calculations we
 find that in the case of discrete energy loss the pile-up, if present,
 is less pronounced, and the cut-off is smoother with particle energies
 extending beyond the cut-off energy.  This may well have
 observational consequences as it will affect the spectrum of
 synchrotron radiation and inverse Compton radiation.

 Further applications of this technique will compare the
 radiation spectra due to synchrotron radiation and to IC energy
 loss by the accelerated electrons and to those that escape from
 the shock in more realistic astrophysical environments, related
 to the environments in different galactic supernova remnants.
 The technique developed can also be used to study the shape of
 the energy spectra of protons accelerated to ultra high energy
 in the presence of energy loss, such as pair production and
 photoproduction on the ambient photon fields.  In addition, the
 Monte Carlo technique developed can be used for studies of the
 shape of the energy spectra of the accelerated particles in the
 presence of energy loss in more complicated acceleration models,
 including non linear shock models where the effective
 compression ratio is a function of the rigidity of the
 accelerated particles. We also intend to apply this method to
 study the shape of the proton spectra accelerated to ultra high
 energy in the presence of pair production and photoproduction
 losses in interactions on the ambient photon fields.

\section*{Acknowledgments}

We thank Troy Porter for helpful suggestions.  T.S. thanks the
University of Adelaide for hospitality during his visit which was
funded partly by the International Visitor Program of the Special
Research centre for Theoretical Astrophysics, University of
Sydney, and partly from an Australian Research Council grant to
R.J.P.  The research of R.J.P. is funded by the Australian
Research Council.  The research of T.S. is funded in part by NASA
grant NAG5-5106.

\newpage

\section*{Appendix A}

Using the Green's function (e.g. \cite{Stecker71}) for the
``leaky-box model'' of cosmic ray propagation, more properly
called the homogeneous model, one can find the steady-state
solution for the energy spectrum of particles inside the
leaky-box
\begin{eqnarray}
{d N_{\rm in} \over dE} = {N_0 \over R(E)} \exp \left\{
- \int_{E_0}^E {d E' \over \tau (E') R(E') } \right\}
\end{eqnarray}
where $R(E)$ is the rate of energy gain and $\tau(E)$ is the
escape time.  We can use this in the present case of the leaky
box acceleration model with, $R(E)=(aE^{1 - \delta} + bE^2)$ and
$\tau(E) = (cE^{-\delta} + cE_{\rm max}^{-\delta})^{-1}$.  We
require the spectrum of accelerated particles, i.e. those
escaping from the leaky-box, and this is simply $d N_{\rm in}/dE$
divided by the escape time $\tau(E)$:
\begin{eqnarray}
{d N \over dE} = {\frac{c\, {E^{-\delta}} + c {E_{\rm max}^{-\delta}} }
       {a\,{E^{1 - \delta}}- b\,{E^2}}} N_0 \exp [ - I(E)]
\end{eqnarray}
where
\begin{eqnarray}
I(E) & \equiv &
\int_{E_0}^E {\frac{c\, {E^{-\delta}} + c {E_{\rm max}^{-\delta}} }
       {a\,{E^{1 - \delta}}- b\,{E^2}}}\,dE,\\
 &=& \left[
\left( {c \over
       a\,\delta} \right) \left( {E \over E_{\rm max}}
       \right)^\delta {_2F_1} \! \left( {\frac{\delta}{1 +
       \delta}},1, 1 + {\frac{\delta}{1 +
       \delta}},{\frac{b\,{E^{1 + \delta}}}{a}} \right) \right. \nonumber
       \\ && \left. - \;\frac{c}{a( 1 + \delta) }\ln \! \left( b - {a \over
       E^{1 + \delta}} \right) \right]_{E_0}^E ,
\end{eqnarray}
and ${_2F_1}$ is the hypergeometric function.

Substituting the integration limits, and using $\Gamma = (1 + c/a)$,
 and $E_{\rm cut}= (a/b)^{1/(1+\delta)}$ we obtain
\begin{eqnarray}
{d N \over dE} = \frac{(\Gamma-1)( E^{-\delta} + E_{\rm max}^{-\delta}) }
       {E^{1 - \delta}- E^2E_{\rm cut}^{-(1 + \delta)}} N_0 \exp [ - I(E)]
\end{eqnarray}
where
\begin{eqnarray}
I(E) &=& \left( {\Gamma - 1 \over \delta} \right) \left( {E \over
	E_{\rm max}} \right)^\delta \left\{ {_2F_1} \! \left[
	{\frac{\delta}{1 + \delta}},1, 1 + {\frac{\delta}{1 +
	\delta}}, \left( {E \over E_{\rm cut}} \right)^{1 +
	\delta} \right] \right. \nonumber \\ && - \left. {_2F_1}
	\! \left[ {\frac{\delta}{1 + \delta}},1, 1 +
	{\frac{\delta}{1 + \delta}}, \left( {E_0 \over E_{\rm
	cut}} \right)^{1 + \delta} \right] \right\} \nonumber \\
	&& - \left( {\Gamma - 1 \over 1 + \delta} \right) 
	\ln \! \left( { E_{\rm cut}^{-(1 + \delta)} 
	- E^{-(1 + \delta)}  \over E_{\rm cut}^{-(1 + \delta)} 
	- E_0^{-(1 + \delta)} } \right).
\end{eqnarray}

\newpage

\end{document}